# A semiconductor photon-sorter

A. J. Bennett[1*], J. P. Lee[1, 2], D. J. P. Ellis[1], I. Farrer[3*], D. A. Ritchie[3], and A. J. Shields[1].

[1]Toshiba Research Europe Limited, Cambridge Research Laboratory, 208 Science Park, Milton Road, Cambridge, CB4 0GZ, U. K.

[2] *Engineering Department, University of Cambridge, 9 J.J. Thomson Avenue, Cambridge, CB3 0FA, U.K.*

[3] *Cavendish Laboratory, Cambridge University, J. J. Thomson Avenue, Cambridge, CB3 0HE, U. K.*

* *Department of Electronic & Electrical Engineering, University of Sheffield, Mappin St, Sheffield, S13JD, U.K.*

Photons do not interact directly with each other, but conditional control of one beam by another can be achieved with non-linear optical media at high field intensities. It is exceedingly difficult to reach such intensities at the single photon level but proposals have been made to obtain effective interactions by scattering photons from single transitions. We report here effective interactions between photons created using a quantum dot weakly coupled to a cavity. We show that a passive single-photon non-linearity can modify the counting statistics of a Poissonian beam, sorting the photons in number. This is used to create strong correlations between detection events and sort polarisation correlated photons from an uncorrelated stream using a single spin. These results pave the way for optical switches operated by single quanta of light.

In many ways photons are ideal carriers of classical and quantum information in that they can be transmitted over large distances and only suffer decoherence from absorption, which engineers can minimize[1]. Conversely, for light to process encoded information it must be coupled to an optically active medium. In the case of quantum optics, non-linear crystals have been developed that can create entangled photon pairs probabilistically[2]. This precludes their operation with single-photon input states; it remains desirable to create a single-photon non-linearity for optical quantum information processing[3,4].

In systems where light and matter are strongly coupled the inherent non-linearity of the Jaynes-Cummings ladder can be used to manipulate photon states[5,6]. It has been shown that the same functionality can be achieved by single emitters coupled to an open optical system, such as a 1D photonic crystal mode or a surface plasmon[7,8], where near deterministic interactions are possible in high efficiency geometries. The promise of this approach has led to many theoretical proposals to create giant Kerr rotations of photon polarisation[9,10] entangle spins and photons[11,12], perform logic operations[13,14,15] and achieve rectification of light transmission[16], to name a few possibilities. Experiments have recently shown the optical transitions in ions can mediate an attractive force between photons, leading to bunching being observed in a transmitted photon stream[17]. It is appealing to realize these applications in the solid-state where devices can be miniaturized and produced at scale.

Figure 1 shows the concept: a waveguide supporting a single transverse optical mode hosts a single transition that controls the transmission and reflection of light. In Fig. 1c we plot the real and imaginary components of the transition's fluorescence. The net optical response of this system is governed by the coherent interference of the fluorescence and the incident light, which we show leads to strong correlations between photons in the reflected beam. The cavity formed by mirrors M1 and M2 enhances the coherent part of the reflected beam, leading to efficient "photon sorting"; measurement of one reflected photon is correlated with others in the same beam even when the incident light displays no correlation.

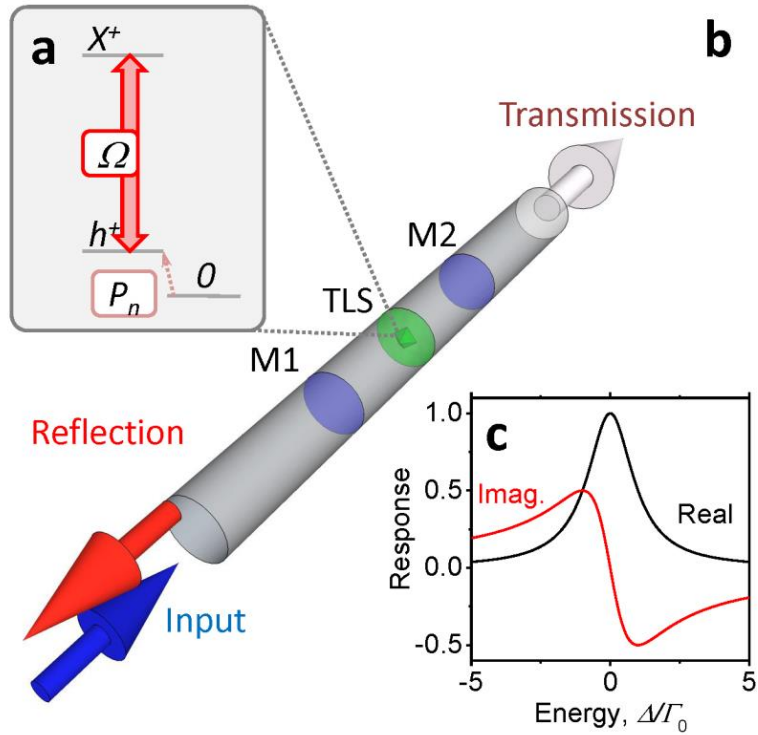

**Figure 1 a**, The level scheme of the emitter. The thick red arrow couples to the transition between and single hole ($h+$) and a positively charged exciton ($X+$) **b**, shows a schematic of system where a single transition (green) can control the transmission and reflection through the guide. **c,** calculated real (black) and imaginary (red) parts of the electric field of the transition as a function of detuning, $\Delta$, from the transition of width $\Gamma_0$.

The experiments reported here used the reflected light from a pillar micro-cavity[18] containing a single InGaAs quantum dot emitting at 1332 meV. The transition between the trion state $X^+$ and a single hole, was activated by a weak non-resonant optical field; effectively $P_n$ in Fig. 1a introduced a hole to the ground state. Lateral confinement led to degenerate $HE_{11}$ optical modes whilst Bragg reflectors confined light along the axis, leading to a Beta factor of 0.9. The Purcell effect further enhanced the fraction of light Resonantly Rayleigh Scattered (RRS) from the transition[19,20,21] to near unity at a Rabi frequency of 0.83 GHz. The RRS photons retain a high spectral overlap with the input optical field, allowing for high level of first-order interference.

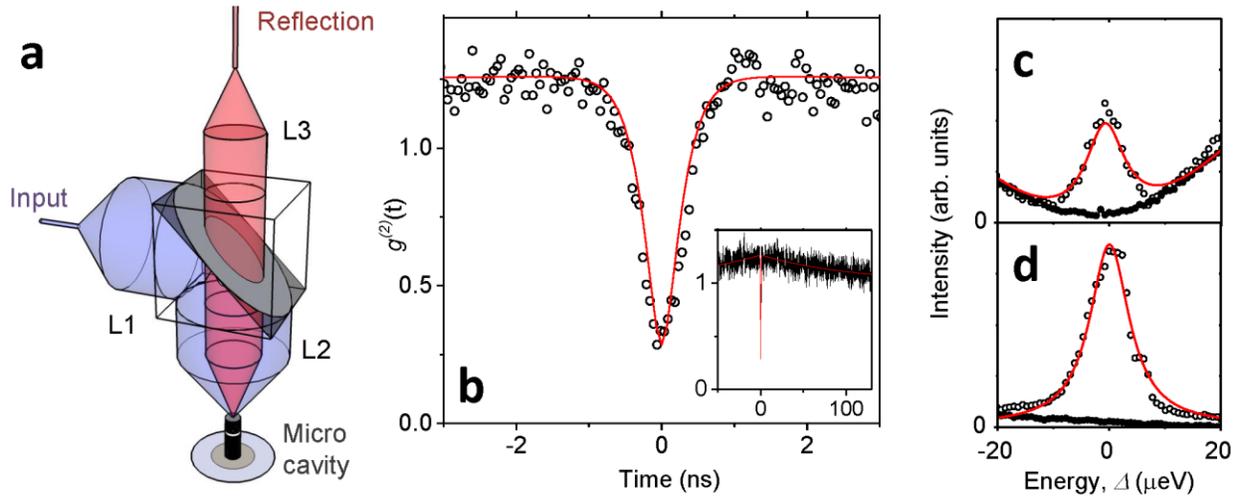

**Figure 2. a**, A schematic of the experimental layout. **b**, The auto-correlation histogram of the light collected without polarisation filtering at 11K. The insert shows data over a wider range to illustrate the charge-induced bunching caused by the probabilistic excitation with laser $P_n$. **c**, shows a scan of intensity as a function of energy detuning with (open data points) and without (filled data points) a charge in the ground state, when the collected polarisation is parallel to that of the driving field. **d**, shows the same scan as in **c**, with the collected polarisation set orthogonal to the driving field.

Fig. 2a shows an input beam focused onto the sample. Light was collected along the same axis, after transmission though a weakly reflecting beam-splitter, into a single mode output fiber. This confocal arrangement with unequal lenses L1 and L3 limits the reflectivity of the cavity to the point where the light in the output fiber is controlled by the transition. A Hanbury-Brown and Twiss autocorrelation histogram confirms the light displays non-classical photon statistics (Fig. 2 b), with $g^{(2)}(0) = 0.28$. Note that there were no polarisation optics in the detection path so, in effect, the Poissonian statistics of the input field were "truncated" by the transition, removing N ≥ 2 photon components on the timescale of the transition lifetime.

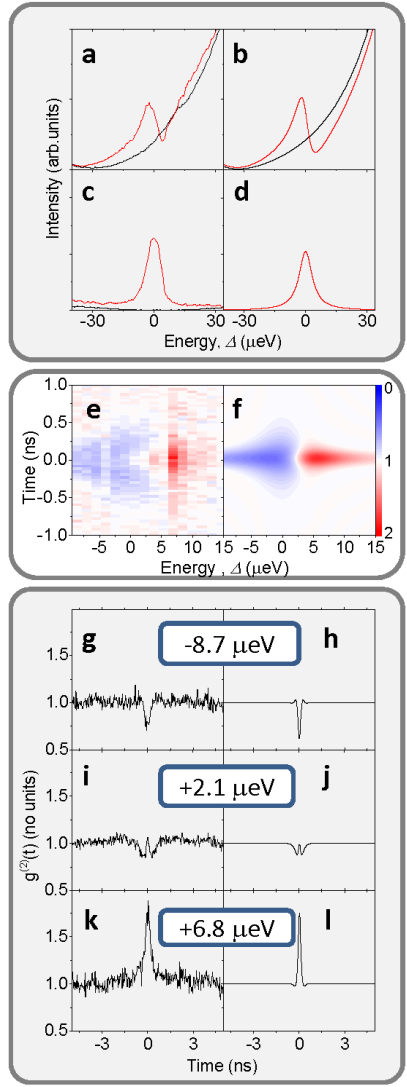

**Figure 3 a,** The collected intensity when the input and detection polarisations are equal, with (red) and without (black) a carrier present in the ground state at 5 K. **b**, Calculation of the same measurement. **c**, and **d**, show experiment and model, respectively, under the same conditions but with the collection polarisation set to reject the light reflected by the cavity. **e,** and **f,** show experiment and model of the photon statistics as a function of detuning. The lower panel shows individual photon auto-correlation measurements at fixed detuning of **g,** and **h,** -8.7μeV, **i,** and **j,** 2.1 μeV and **k,** and **l,** +6.8 μeV.

Introducing a polariser in the detection path set parallel to the driving field polarisation the RRS intensity was an order of magnitude greater than the detected light reflected by the cavity. The reflectivity spectrum as a function of energy detuning, $\Delta$, is shown in Fig. 2c when the transition was active (open circles). When the ground state was not loaded with a hole the reflectivity spectrum shows the response of the cavity mode (filled circles). Conversely, when the polariser is set to be orthogonal to the driving field we observe the real-part of the transition's response (Fig. 2d, open circles). Following analysis, we conclude the non-zero $g^{(2)}(0)$ measured in Fig. 2b is a result of the mixing ideal anti-bunched RRS photons with the small contribution of Poissonian light reflected directly by the cavity.

The imaginary part of the transition's fluorescence can be revealed by mixing the RRS photons with a phase-coherent reference beam. Fig. 3a shows a measurement of the net reflectivity for co-polarised arrangement when the transition was temperature-shifted to a point of greater cavity reflectivity. The electric field at the output, $E_{total}(\Delta)$, is given by[22,23]:

$$E_{total}(\Delta) = \sqrt{R_{cavity}(\Delta)} + \frac{\Gamma_0 \sqrt{R_0}}{\Delta + i\Gamma_0}$$

(1)

Where $R_0$ is the maximum reflectivity of the transition and $R_{cavity}(\Delta)$ the cavity reflectivity determined from the experimental data in Fig. 3a (black line). The reflection from the dot is a Lorentzian function with full-width $\Gamma_0 = 7.7$ µeV (Fig. 3c) cavity-enhanced by the Purcell effect. The calculated reflectivity is given by $|E_{total}|^2$, which is shown in Fig. 3b, shows a good agreement with the experiment. We see that the peak reflectivity at negative $\Delta$ is increased by 210% relative to the case where the transition is inactive, and is suppressed by 26% for positive $\Delta$. This 236% modulation in the signal can be tuned with temperature and input intensity, and underlines the degree of control the transition has over the reflected light.

As the input energy was scanned across the transition the photon statistics at the detector were modified as shown in Fig. 3e, and modelled in Fig. 3f. These auto-correlations can be understood by combining the interference effect described by equation (1) and the dynamics of the transition after a photon detection event. Intuitively, this effect is a result of the quantized RRS photons being added or subtracted from the output-field. When the input is blue-detuned from the transition a single photon is added in phase with the coherent light and so the intensity is increased, but the corresponding $g^{(2)}(0)$ parameter is then less than one. At -8.7 µeV $g^{(2)}(0) =$ 0.75 (Fig. 3g), in agreement with the model Fig. 3h. As $\Delta$ is increased it is possible to observe some bunching around time zero. At $\Delta = +2.1$ µeV (Fig 3i) a W-shaped correlation observed and predicted. When the input was red detuned the RRS light was subtracted from the coherent field, reducing the total intensity but leaving the multi-photon components of the remaining state to dominate. At $\Delta = +6.8$ µeV $g^{(2)}(0)=1.75$. The ability of the transition to actively sort the input weak coherent beam by photon number has led to strong correlations in the reflected photon statistics. This gives the appearance of an attractive or repulsive force either bunching or separating single photons from each other.

Finally, we used the spin of the hole to sort the photons in polarisation, under the same conditions as in Fig. 2b. Light at the detector was dominated by RRS photons, so the detection of a certain handedness of photon corresponded uniquely to a certain spin-eigenstate. Polarised correlations measure the probability of the source scattering incident linearly polarised photons into the same circular polarisation, *left-left* (L-L) or *right-right* (R-R), versus the probability of scattering photons into consecutive orthogonal circular polarisations L-R or R-L. Only when the hole spin is aligned along the cavity axis can it result in polarisation correlation at the detectors. When the spin is not prepared its random orientation and zero expectation value lead to a 1/3 degree of polarisation correlation, as is observed in Fig. 4b. The bunching of photons of common circular polarisation has a characteristic timescale of 0.315 ± 0.003 µs at zero magnetic field. This value is limited by the incoherent, probabilistic charging of the dot but could be increased to >100 µs by deterministic charging[24].

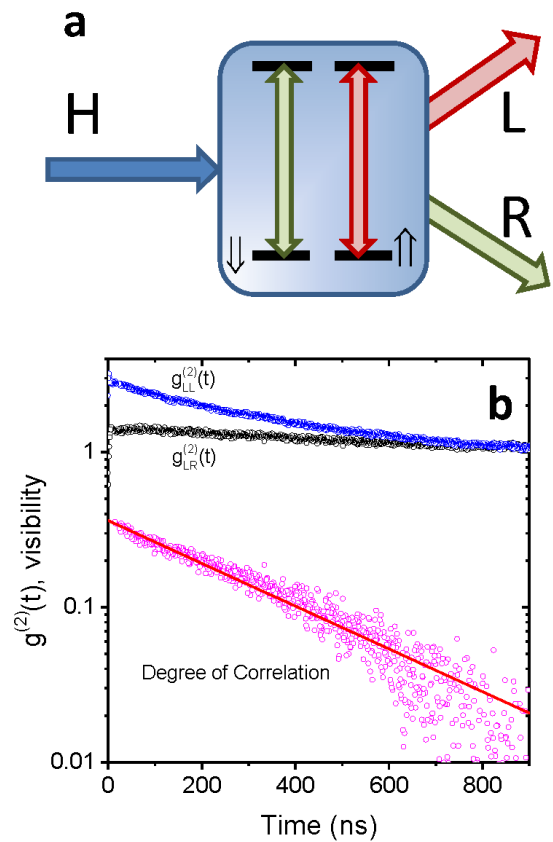

**Figure 4 a,** A linearly polarised input is scattered from the two degenerate transitions, each coupled equally to a different ground state hole-spin. The scattered photon polarisation is determined by this spin. **b,** Polarised correlation measurements on the scattered photons, for (blue) common circular polarisation and (black) orthogonal circular polarisation and the degree of polarisation correlation (pink).

The experiment reported here show a transition is able to sort photons in a weak coherent beam based on photon number and polarisation. The coherently scattered light from the transition passively creates a non-linear interaction between photons that can be used to create strongly correlated photons without requiring strong coupling or intense optical fields. This photon sorter will operate with any incident beam which is at the energy of the transition provided its spectral width is less than that of the transition, making it compatible with atomic systems and narrow-band quantum memories.

We anticipate that increasing the vertical extent of the micro-cavity, whilst ensuring the emitter remains in the confocal image plane will further reduce the cavity's reflection, and may allow for the inclusion of electrical contacts to control the phase shift imparted by the transition[25]. Furthermore, optimizing the overlap of the input beam with the transition could lead to their deterministic interaction[26,27] leading to a single-photon transistor. Most promisingly, extension of this work to a system where the spin of the hole can be optically prepared[28] will lead to an efficient photon-photon or spin-photon entanglement resource[11,12,13,14,15].